\documentclass[12pt]{iopart}

\usepackage{graphicx}
\usepackage{iopams}
\begin{document}

\title{Static and dynamic heterogeneities in irreversible gels and colloidal gelation
}
\author{A. Coniglio, T. Abete, A. de Candia, E. Del Gado, A. Fierro}
\address{Dipartimento di Scienze
Fisiche, Universit\`{a} degli Studi di Napoli ``Federico II'', INFN,
CNR-INFM  Coherentia and CNISM, via Cinthia, 80126 Napoli, Italy}
\begin{abstract}
We compare the slow dynamics of irreversible gels, colloidal gels,
glasses and spin glasses by analyzing the behavior of the so called
non-linear dynamical susceptibility, a quantity usually introduced
to quantitatively characterize the dynamical heterogeneities. In
glasses this quantity typically grows with the time, reaches a maximum
and then decreases at large time, due to the transient nature of
dynamical heterogeneities and to the absence of a diverging static
correlation length. We have recently shown that in irreversible gels
the dynamical susceptibility is instead an increasing function of
the time, as in the case of spin glasses, and tends asymptotically
to the mean cluster size. On the basis of molecular dynamics
simulations, we here show that in colloidal gelation where clusters
are not permanent, at very low temperature and volume fractions,
i.e. when the lifetime of the bonds is much larger than the
structural relaxation time, the non-linear susceptibility has a
behavior similar to the one of the irreversible gel, followed, at
higher volume fractions, by a crossover towards the behavior of glass forming liquids.
\end{abstract}
\date{\today}
\maketitle

\section{Introduction}
Many complex systems, such as glasses, spin glasses, irreversible
gels, colloidal gels and others, exhibit a complex dynamics, all
characterized by a slowing down usually leading to a structural
arrest. Nevertheless there are significative differences, among
these systems, often not well clarified. One of the key concept to
describe the slow dynamics in glassy systems is the concept of
dynamical heterogeneities. Here we want to classify and compare the
above systems by looking at the behavior of the so called non-linear
dynamical susceptibility, a quantity usually introduced to
quantitatively characterize the dynamical heterogeneities. In glass
forming liquids different definitions have been proposed for the non
linear dynamical susceptibility \cite{franz, biroli}. Mostly
considered are the fluctuations of the self and the total
intermediate scattering functions. These quantities typically have
similar behavior, grow with the time, reach a maximum and then
decrease at large time. This behavior is a consequence of the
transient nature of dynamical heterogeneities and the absence of a
diverging static correlation length. On the other hand for systems
with quenched interactions such as spin glasses, characterized by a
diverging static correlation length at spin glass critical
temperature, the dynamical susceptibility defined as the
fluctuations of the time dependent spin-spin autocorrelation
function is a monotonic function increasing with the time. As the
time goes to infinity this dynamical susceptibility for a fixed
temperature T tends to a plateau whose value coincides with the
static non-linear susceptibility. Therefore as T approaches the spin
glass temperature the value of the plateau diverges as the static
non-linear susceptibility. In irreversible gels the definition of
the corresponding dynamic susceptibility is not straightforward and
one should carefully distinguish different dynamical quantities. We
have recently shown \cite{tiziana_prl} that in a microscopic model
for irreversible gels the dynamical susceptibility defined as the
fluctuations of the self-intermediate scattering function is a
monotonic function as in the case of spin glasses and, for each
fixed value of the volume fraction, its long time limit tends to a
plateau whose value coincides with the mean cluster size. The value
of this plateau therefore diverges at the percolation threshold as
the mean cluster size. Such finding corresponds to the fact that in
irreversible gelation the heterogeneities are due to the static
nature of the clusters.

On this basis, we speculate that in colloidal gelation, where
clusters are not permanent due to the finite bond lifetime, this
non-linear susceptibility should show a behavior similar to the
dynamical susceptibility of the irreversible gel at very low
temperature and very low volume fraction, where the lifetime of the
bonds is much larger than the structural relaxation time. At higher
volume fractions and temperature, it should crossover towards a
behavior of glass forming liquids. Here we give some evidence
based on some molecular dynamics simulations of a model for
colloidal gelation \cite{physicaa,tubi,dlvo_prep}. Moreover, using
this scenario, we interpret previous results found in experimental
investigations of colloidal suspension \cite{cipelletti} and in some
molecular dynamics simulations \cite{physicaa}. Finally, we show
that also in spin glass type of models when the lifetime of the
interaction is made finite the behavior found is similar to that
found in colloidal systems \cite{fdc}. 

In the following, first we
recall the behaviour of linear dynamical susceptibility of glass
forming liquids and spin glasses (Sect.\ref{sglass}). Later
(Sect.\ref{gels}) we consider the case of irreversible gels. In
Sect.\ref{collgel}, we discuss the case of colloidal gelation and
compare with a spin glass type of model with annealed interactions
(Sect\ref{ann}). Finally we analyze the emerging scenario and the
further developments that this study suggests (Sect.\ref{conclu}).

\section{Systems with quenched interactions}
\subsection{Spin glasses}
\label{sglass}
We briefly recall the $3d$ Ising spin glass where the Hamiltonian of the model
is $H = J\sum_{\langle ij \rangle} \epsilon_{ij}S_i S_j$, with $S_i=\pm 1$
Ising spins,
and $\epsilon_{ij}=\pm 1$ quenched and disordered interactions. The $3d$
model undergoes a transition at a temperature, $T_{SG}$,
with a divergence of the static non-linear susceptibility,
$\chi_{nl}={1\over N}\sum_{ij} [\langle S_i S_j\rangle^2]$,
where the average $\langle\cdots\rangle$ is over the Boltzmann
measure, and the average $[\cdots]$ is over the disorder configurations.
The dynamical non-linear susceptibility was firstly introduced in
$p$-spin models, considered as the prototype models of glass formers in mean
field \cite{franz},
\begin{equation}
\chi(t)=N[\langle q(t)^2\rangle-\langle q(t)\rangle^2],
\end{equation}
where $q(t)={1\over N}\sum_i  S_i(t')S_i(t'+t)$ and
the average $\langle\cdots\rangle$ is done on the reference time $t'$.
In the $3d$ Ising spin glass, differently from the behavior observed in $p$-spin models,
$\chi(t)$ grows monotonically until a plateau value is reached. The
plateau value coincides with the static non-linear susceptibility and diverges
at the transition \cite{fdc}.

\subsection{Irreversible gels}
\label{gels}
In this section we present a molecular dynamics (MD) study of a microscopic
model recently introduced \cite{tiziana_prl} for irreversible gels.
We consider a $3d$ system of $N=1000$ particles
interacting via a Lennard-Jones potential, truncated in order to
have only the repulsive part:
$$
U_{ij}^{LJ}=\left\{ \begin{array}{ll}
4\epsilon[(\sigma/r_{ij})^{12}-(\sigma/r_{ij})^6+\frac{1}{4}], & r_{ij}<2^{1/6}\sigma \\
0, & r_{ij}\ge2^{1/6}\sigma \end{array} \right.
$$
where $r_{ij}$ is the distance between the particles $i$ and $j$.
After a first equilibration, we introduce quenched bonds between particles
whose relative distance is smaller than $R_0$ by adding
an attractive potential:
$$
U_{ij}^{FENE}=\left\{ \begin{array}{ll}
-0.5 k_0 R_0^2 \ln[1-(r_{ij}/R_0)^2], & r_{ij}< R_0\\
\infty, & r_{ij}\ge R_0 \end{array} \right.
$$
representing a finitely extendable non-linear elastic
(FENE) \cite{FENE}. The system is then further thermalized. We have
chosen $k_0=30\epsilon/\sigma^2$ and $R_0=1.5\sigma$ as in Ref.
\cite{FENE} and performed MD simulations in a box of
linear size $L$
(in units of $\sigma$) with periodic boundary conditions. The
equations of motion were solved in the canonical ensemble (with a
Nos\'e-Hoover thermostat) using the velocity-Verlet algorithm
\cite{Nose-Hoover} with a time step $\Delta t=0.001\delta\tau$,
where $\delta\tau=\sigma(m/\epsilon)^{1/2}$, with $m$ the mass of
particle. In our reduced units the unit length is $\sigma$, the unit
energy $\epsilon$ and the Boltzmann constant $k_B$ is set equal to
$1$. The temperature is fixed at $T=2$ and the volume fraction
$\phi=\pi\sigma^3N/6L^3$ is varied from $\phi=0.02$ to $\phi=0.2$.
By varying the volume fraction we find that the system undergoes a
percolation transition  in the universality class of the
random percolation.
We have investigated the dynamical evolution of the system by studying
the self intermediate scattering functions,
$\Phi_s(k,t)=\frac{1}{N}\sum_{i=1}^N
e^{i\vec{k}\cdot(\vec{r}_i(t)- \vec{r}_i(0))}$,
for increasing volume fractions. As in experiments on real gels, we observe
stretched exponential decays at volume fraction lower than the percolation
threshold  $\phi_c$, and approaching $\phi_c$ the onset of power law decays.

We have moreover measured the dynamical susceptibility
associated to the fluctuations of the self intermediate scattering functions
\cite{biroli},
i.e. $\chi_4(k,t)=N\left[\rule{0pt}{10pt}\langle
|\Phi_s(k,t)|^2\rangle-\langle \Phi_s(k,t)\rangle^2\right]$, where
$\langle \dots \rangle$ is the thermal average for a fixed bond configuration
and $[\dots]$ is the average over the bond configurations.
We have shown analytically in Ref. \cite{tiziana_prl} that this quantity, in the
thermodynamic limit, for $t\rightarrow\infty$  and $k\rightarrow 0$, tends
to the mean cluster size.

\begin{figure}
\begin{center}
\includegraphics[width=8cm]{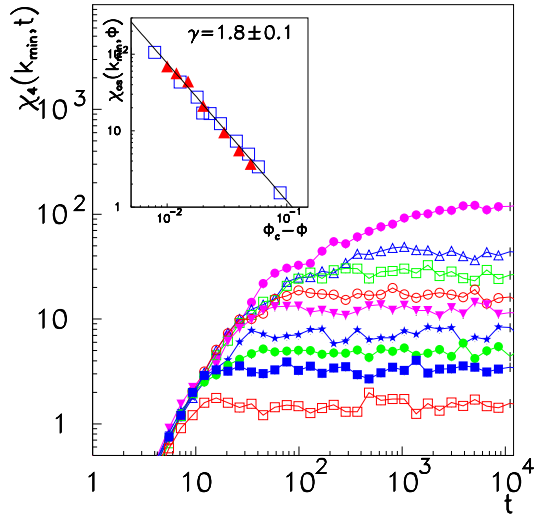}
\caption{(Color online){\bf Main frame}: $\chi_4(k_{min},t)$ as a function of $t$ for
$\phi= 0.02$, $0.05$, $0.06$, $0.07$, $0.08$, $0.085$,
$0.09$, $0.095$, $0.10$ (from bottom to top). {\bf Inset}:
Asymptotic values of the susceptibility (full triangles),
$\chi_{as}(k_{min},\phi)$ and mean cluster size (open
squares) as a function of $(\phi_c-\phi)$.
The data are fitted by the power law
$(\phi_c-\phi)^{-\gamma}$ with $\gamma=1.8\pm 0.1$.}
\label{fig5}
\end{center}
\vspace{-0.5cm}
\end{figure}

In the Main frame of Fig.\ref{fig5}, $\chi_4(k_{min},t)$ 
(with $k_{min}=2\pi/L$) is plotted
for increasing values of the volume fractions $\phi \le
\phi_c$.
Differently from the non monotonic behavior typically
observed in glassy systems, we find that it
increases with time until it reaches a plateau,
whose value increases as a function of $\phi$.

In the Inset of Fig.\ref{fig5}, $\chi_{as}(k_{min},\phi)\equiv
\lim_{t\to\infty} \chi_4(k_{min},t)$
is plotted as a function of $(\phi_c-\phi)$ together with the mean cluster size.
We find that, as the percolation threshold is approached from below,
$\chi_{as}(k_{min},\phi)$ diverges as a power law at $\phi_c$.
The exponent, within the numerical accuracy, is in agreement
with the value of the exponent $\gamma$ of the mean cluster size.
This finding confirms that
one key difference between irreversible gelation due to chemical bonds and
supercooled liquids close to the glass transition is that in
irreversible gelation the heterogeneities have a static nature
(clusters). The clusters, on the other hand, affect the dynamics
and as a consequence the dynamic transition coincides with the
static transition, characterized by the divergence of a static
correlation length (linear size of the clusters).

The behavior observed in the case of permanent gels is very
similar to that of spin glasses in finite dimensions. Although these systems
have a very different structures, they show a very similar dynamical behavior
due to the static nature of heterogeneities.
In both cases interactions are
quenched.
What can we expect in the cases where the interactions are not
quenched and have a finite lifetime? In the following section we will try to
answer to this question.

\section{Systems with finite bond lifetime}
\subsection{Colloidal gelation}
\label{collgel}
In this section we present the results obtained in a MD study of a DLVO-type
potential \cite{dlvo} for charged colloidal systems, and discuss how in
colloidal gelation the finite bond lifetime affects the dynamics and, in
particular, the behavior of the dynamical susceptibility.

We consider a system of $N=10000\,\phi$ particles, interacting via a DLVO-type
potential, which contains a Van der Waals type interaction plus an effective
repulsion due to the presence of charges:
\begin{equation}
V(r)=\epsilon \left[a_1 \left(\frac{\sigma}{r}\right)^{36}
-a_2\left(\frac{\sigma}{r}\right)^6+a_3e^{-\lambda(\frac{r}{\sigma}-1)}\right],
\label{potential}
\end{equation}
where $a_1=2.3$, $a_2=6$, $a_3=3.5$, and $\lambda =2.5$.
With these parameters
the repulsion term dominates the Van der Waals attraction at long
range, providing a short range attraction and a long range repulsive barrier.
The potential is truncated and shifted at a distance of 3.5$\sigma$.
To mimic the colloidal dynamics, we performed
MD simulations at constant temperature.
Equations of motion were solved in the canonical ensemble (with a
Nos\'e-Hoover thermostat) using a velocity Verlet algorithm
with a time step of $0.001 t_0$ (where $t_0=\sqrt{\frac{m
\sigma^2}{\epsilon}}$ and $m$ is the mass of the particles).

We find \cite{physicaa, tubi, dlvo_prep} that at low volume fraction compact
stable clusters form with typical size $s\simeq10$. By increasing the volume
fraction a residual attractive interaction between the clusters produces
elongated structures, which finally order in a columnar phase \cite{tubi}.
A small degree of polidispersity is introduced \cite{dlvo_prep} in order to
avoid the transition to the ordered phase.  In this case, by increasing the
volume fraction the elongated structures form instead
a long-living random percolating network, i.e. the gel phase.
The bond lifetime has a non monotonic behaviour: At $k_B T=0.15\epsilon$ it
decreases of about one order of magnitude from $\phi=0.10$ to $0.13$, has
a minimum at $0.13$, and finally increases for $\phi>0.13$.

\begin{figure}
\begin{center}
\includegraphics[width=8cm]{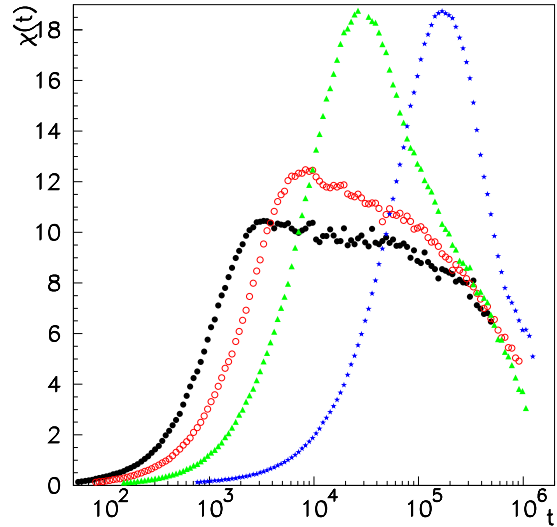}
\caption{(Color online) The dynamic susceptibility, $\chi_4(k_{min},t)$, for
$k_B T=0.15\epsilon$ and
$\phi=~0.10$, $0.11$, $0.12$, $0.13$ (from left to right).}
\label{figure5}
\end{center}
\vspace{-0.5cm}
\end{figure}

The dynamical susceptibility, $\chi_4(k_{min},t)$,
is measured for $\phi=0.10$, $0.11$, $0.12$,
$0.13$ (see Fig.\ \ref{figure5}). The data give evidence of a clear crossover
from the low volume fraction regime to the intermediate regime. In
the first regime, where the bond lifetime is much
larger than the
structural relaxation time, we find a behavior resembling that observed in
the permanent gels: Although $\chi_4(k_{min},t)$ is a non monotonic function,
it increases until a value comparable to the mean cluster size is reached;
A plateau decreasing slowly as a
function of time is clearly present in the intermediate time region;
Finally, at very long times, $\chi_4(k_{min},t)$ decreases to its equilibrium
value. Increasing $\phi$, the bond lifetime and the structural relaxation time
become comparable: In this case $\chi_4(k_{min},t)$, which  is again a non
monotonic function,
displays a well pronounced maximum as usually observed in glassy systems.
This data suggest that in the first regime the clusters
behave dynamically as made of permanent bonds as in chemical gelation, and the dynamics
is dominated by the presence of such clusters. Increasing
$\phi$, the structural relaxation begins to be affected also by the crowding of the
particles, and a clear crossover to a new glassy regime is found.

\subsection{A spin glass type of model with annealed interactions}
\label{ann}
It is now extremely interesting to analyse the case of annealed interactions
in a spin glass type of model.
To this aim we consider the results obtained in Monte Carlo
simulations of the so called frustrated lattice gas (FLG). This model, recently
introduced in connection with the glass transition \cite{varenna,meanfield},
has mean field properties closely
related to those of $p$-spin models. Being constituted
by diffusing particles, it is suited to study quantities like the diffusion
coefficient, or the density autocorrelation functions, that are usually
important in the study of liquids. The Hamiltonian of the model is:
\begin{equation}
-\beta H = J\sum_{\langle ij \rangle}
(\epsilon_{ij}S_i S_j - 1)n_in_j +\mu \sum_i n_i,
\label{flg}
\end{equation}
where $S_i=\pm 1$ are Ising spins, $n_i=0,1$ are occupation variables,
and $\epsilon_{ij}=\pm 1$.
In the case where $\epsilon_{ij}$ are quenched variables randomly distributed
the $3d$ model undergoes, a transition of the type of $3d$ spin glasses
\cite{antonio}.
Here we considered the case where the interactions $\epsilon_{ij}$ evolve in
time, i.e. they are annealed variables \cite{fdc}.
In this case dynamical properties strongly resembling those of glass formers
and well fitted by the mode coupling theory for supercooled liquids are found.

\begin{figure}
\begin{center}
\includegraphics[width=8cm]{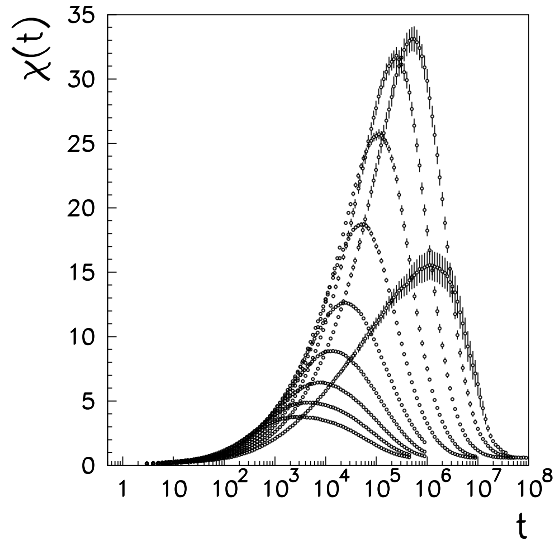}
\caption{Dynamical non-linear susceptibility in the annealed FLG for densities
$\rho=0.52$, $0.53$, $0.54$, $0.55$, $0.56$, $0.57$, $0.58$,
$0.59$, $0.60$, $0.61$ (from left to right).}
\label{fig_chi}
\end{center}
\end{figure}
The dynamical non-linear susceptibility is defined by
\begin{equation}
\chi(t)=N[\langle q(t)^2\rangle-\langle q(t)\rangle^2],
\label{eq_chi}
\end{equation}
where $q(t)={1\over N}\sum_i  S_i(t')n_i(t')S_i(t'+t)n_i(t'+t)$ and
the average $\langle\cdots\rangle$ is done on the reference time
$t'$. In Fig. \ref{fig_chi}, $\chi(t)$ is plotted for increasing
values of the density. The same behavior of the $p$-spin model in
mean field \cite{franz} and of MD simulations of the Lennard-Jones
binary mixture \cite{glotzer} is observed: $\chi(t)$ shows a
maximum, $\chi(t^*)$, that seems to diverge together with the time
of the maximum $t^*$, when the density grows. For the highest
density, the maximum of $\chi(t)$ decreases, possibly due to the
transition to an unfrustrated state.
Comparing
the behavior found in this case with that shown in the previous
section, we note that here the first regime with a clear plateau in
the susceptibility is not present. This is probably due to the fact
that the interaction relaxation time is in this case always
comparable to the structural relaxation time.

\section{Conclusions}
\label{conclu}
By means of the dynamical susceptibility, we have here analyzed the presence of dynamical
heterogeneities in systems with quenched and annealed interactions.
In the case of quenched interactions, as it happens in spin glasses, the
dynamical
susceptibility grows monotonically in time until a plateau value is reached. The
plateau value coincides with the static non-linear susceptibility and diverges
at the transition. This behavior is in fact also observed in a microscopic
model for irreversible gels, where the plateau value of the dynamic
susceptibility diverges at the percolation transition as the mean cluster
size.
These results confirm that in irreversible gelation the heterogeneities have a static nature
(clusters). These clusters, on the other hand, affect the dynamics and as a consequence the
dynamic transition coincides with the static transition, characterized by the divergence
of a static correlation length (linear size of the clusters).

With annealed interactions instead, in the case of spin glass type
of models, one recovers the non monotonic behavior of the dynamical
susceptibility, which is typically observed in glasses. This is due
to the transient nature of dynamical heterogeneities and the absence
of a diverging static correlation length. We analyze moreover the
case of colloidal gelation, where the clusters are not permanent due
to the finite lifetime of the bonds. We find that the dynamical
susceptibility is again a non monotonic function, and displays at
high volume fraction a well pronounced maximum as usually observed in
glassy systems. Remarkably, at very low temperature and very low
volume fraction, where the lifetime of the bonds is much larger than
the structural relaxation time, the dynamical susceptibility shows a
behavior similar to the dynamical susceptibility of the irreversible
gel with a crossover, at higher volume fractions, towards a behavior
typical of glass forming liquids. These results suggest that in the
first regime the dynamics is dominated by clusters, made of bonds
which can be considered as permanent in this time window. Increasing
$\phi$, when these two time scales become comparable, the structural
relaxation begins to be affected also by the crowding of the
particles, and a clear crossover to a new glassy regime is found.

This work has been partially supported by the Marie Curie Fellowship
HPMF-CI2002-01945 and Reintegration Grant MERG-CT-2004-012867,
EU Network Number MRTN-CT-2003-504712, MIUR-PRIN 2004, MIUR-FIRB 2001,
CRdC-AMRA, INFM-PCI.

\end{document}